\documentclass[10pt]{iopart}
\usepackage{iopams}  

\usepackage{graphicx}  

\def\msun{\,{\rm M_\odot}}

\newcommand\beq{\begin{equation}}
\newcommand\eeq{\end{equation}}

\begin{document}

\title[Low-Frequency Gravitational 
Radiation from Coalescing Massive Black Holes]{Low-Frequency Gravitational 
Radiation from Coalescing Massive Black Holes}

\author{A Sesana\dag,
\ F Haardt\dag,\ P Madau\ddag\ and 
M Volonteri\ddag
}

\address{\dag\ Dipartimento di Fisica e Matematica, Universit\`a dell'Insubria,
via Valleggio 11, 22100 Como, Italy}

\address{\ddag\ Department of Astronomy $\&$ Astrophysics, University of
California, 1156 High Street, Santa Cruz, CA 95064}

\begin{abstract}
We compute the expected low-frequency gravitational wave signal from 
coalescing massive black hole (MBH) binaries at the center of galaxies. 
We follow the merging history of halos and associated holes
via cosmological Monte Carlo realizations of the merger hierarchy
from early times to the present in a $\Lambda$CDM cosmology.
MBHs get incorporated through a series of mergers
into larger and larger halos, sink to the centre owing to
dynamical friction, accrete a fraction of the gas in the merger
remnant to become more massive, and form a binary system. Stellar
dynamical processes dominates the orbital evolution of the binary
at large separations, while gravitational wave emission takes over
at small radii, causing the final coalescence  of the system.
We discuss the observability of inspiraling MBH binaries by a low-frequency 
gravitational wave experiment such as the planned {\it Laser Interferometer 
Space Antenna} ({\it LISA}), discriminating between resolvable sources and
unresolved confusion noise. Over a 3-year observing period {\it LISA} 
should resolve this GWB into discrete sources, detecting $\approx 90$ 
individual events above a $S/N=5$ confidence level, while expected confusion
noise is well below planned {\it LISA} capabilities. 
\end{abstract}

%

\section{Introduction}

Studies of gravitational wave (GW) emission and of its detectability are 
becoming increasingly topical in astrophysics. Technological developments of 
bars and interferometers bring the promise of a future direct observation of 
gravitational radiation, allowing to test one of the most fascinating 
predictions of General Relativity, and, at the same time, providing a new 
powerful tool in the astronomical investigation of highly relativistic 
catastrophic events, such as the merging of compact binary systems and the 
collapse of massive stellar cores. Massive black hole  
binaries (MBHBs) are among the primary candidate sources of 
GWs at mHz frequencies ~\cite{enelt, jaffe, uaiti}, the range to be probed 
by the space-based {\it Laser 
Interferometer Space Antenna} ({\it LISA}). Today, massive black holes (MBHs) are ubiquitous in the 
nuclei of nearby galaxies ~\cite{mago}. If MBHs were 
also common in the past (as implied by the notion that many distant galaxies 
harbor active nuclei for a short period of their life), and if their host 
galaxies experience multiple mergers during their lifetime, as dictated by 
popular cold dark matter (CDM) hierarchical cosmologies, then MBHBs 
will inevitably form in large numbers during cosmic history. MBHBs that 
are able to coalesce in less than a Hubble time will give origin to the 
loudest GW events in the universe.

\section {Dynamical evolution of MBHBs}

   \begin{figure}
   \centering
   \includegraphics[width=3.5in]{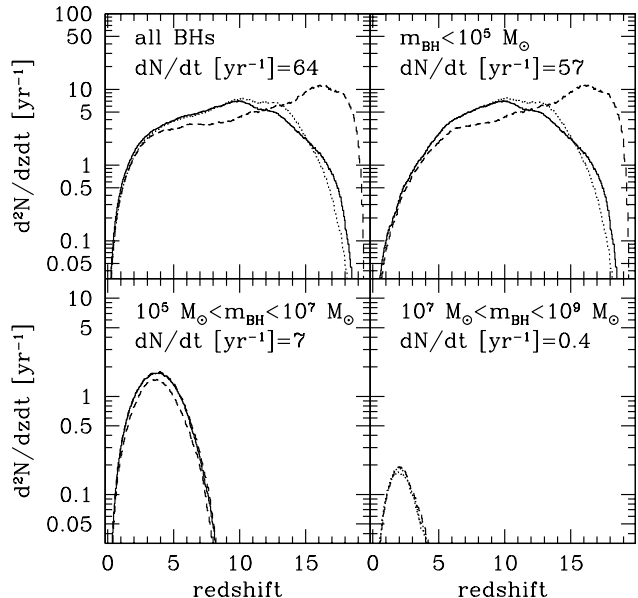}
      \caption{Number of MBHBs 
coalescences observed per year at $z=0$, per unit redshift, in different 
$m_{\rm BH}=m_1+m_2$ mass intervals. Each panel also lists the 
integrated event rate, $dN/dt$, predicted by our model. The rate ({\it 
solid lines}) is  
compared to a case in which triple black hole interactions are switched off 
({\it dotted lines}) and to a case in which  
the hardening of the binary is instantaneous, i.e., MBHBs coalesce 
after a dynamical friction timescale ({\it dashed lines}).
             }
         \label{Fig1}
   \end{figure}

Here we describe the expected GW signal
from inspiraling binaries in a hierarchical structure formation scenario 
in which seed holes of intermediate mass form far up in the dark halo 
``merger tree''.
The model has been discussed in detail in ~\cite{volo}. Seed holes
with $m_{\textrm{seed}}=
150\msun$ are placed within rare 
high-density peaks (minihalos) above the cosmological Jeans and cooling 
masses at redshift 20. Their evolution and growth is followed through Monte 
Carlo realizations of the merger hierarchy, combined with semi-analytical
prescriptions for the main processes involved, such as the dynamical friction
against the dark matter background, the hardening of MBHBs via 
three-body interactions, the MBHBs coalescence due to the emission of 
gravitational waves, triple MBH interactions, and the 
``gravitational rocket'' effect. Quasar activity is triggered during major mergers.
Fig.\ref{Fig1} shows the number of 
MBHB coalescences per unit redshift per unit observed 
year predicted by our model. We expect 
a few tenths of events per year, the vast majority involving quite light
binaries ($m_{\textrm{BH}}=m_1+m_2\leq10^5\msun$, where $m_1$ is the mass of the 
heavier BH in the binary). 

The model reproduces fairly well the observed luminosity function 
of optically-selected quasars in 
the redshift range $1<z<5$, and provides a quantitative explanation 
to the stellar density profiles observed in the cores of bright ellipticals ~\cite{volo2}. 

\section {Gravitational Wave Signal}

   \begin{figure}
   \centering
   \includegraphics[width=3.5in]{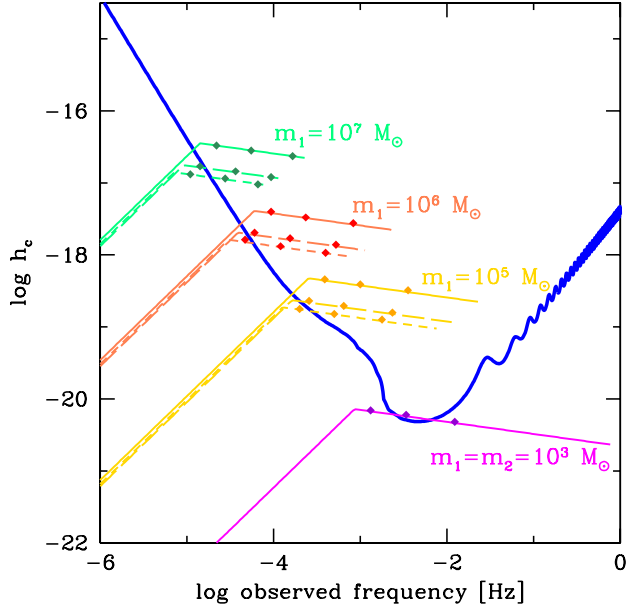}
      \caption{ Characteristic strain spectrum $h_c$ against frequency:
from top to bottom, the three sets of curves refer to systems 
with $\log(m_1/M_{\odot})=7,6,5$ respectively, and the solid, long-dashed,
and short-dashed lines assume the binary at $z=1,3,5$, respectively. 
A mass ratio $m_2/m_1=0.1$ is assumed. The lowest curve is for an equal 
mass binary $m_1=m_2=10^3 M_{\odot}$ at $z=7$. The small 
diamonds on each curve marks, from left to right, 
the frequency 1 year, 1 month and 1 day before coalescence.  
{\it Thick solid curve}: {\it LISA} sensitivity threshold approximatively
accounting for detection with $S/N>5$ including galactic 
~\cite{nelemans} and extragalactic ~\cite{farmer} 
white dwarf binary (WD-WD) confusion noises (added in quadrature). 
A 3-year observation is considered.
              }
         \label{Fig2}
   \end{figure}

We compute the GW emission from each binary in the quadrupole approximation, then, 
to evaluate the total GWB, we integrate the contributions from all binaries ~\cite{s1}.\\
The characteristic amplitude $h_c$ of a source at comoving coordinate distance 
$r(z)$ is 
\beq
h_c \simeq h \sqrt{n}. 
\label{hc}
\eeq
The strain amplitude $h$ (sky-and-polarization averaged) is given by
\beq
h={8\pi^{2/3}\over 10^{1/2}}{G^{5/3}{\cal M}^{5/3}\over c^4 r(z)}f_r^{2/3}, 
\label{eqthorne}
\eeq
where the the rest-frame frequency $f_r$ is related to the {\it observed} 
frequency $f$ as  $f=f_r/(1+z)$, and 
${\cal M}=m_1^{3/5}\,m_2^{3/5}/(m_1+m_2)^{1/5}$ 
is the \emph{chirp mass} of the system.
Here $n$ is the number of cycles a source spends at 
frequency $f_r$, i.e., $n=f_r^2/\dot f_r$ ~\cite{kiptorn}.
Note that for a {\it finite} observation time $\tau$, the number of 
cycles at a given frequency 
$f_r$ can not exceed $f\tau$. 
The behavior of $h_c$ vs. $f$ is shown, for different binaries, in fig. \ref{Fig2}.

   \begin{figure}
   \centering
   \resizebox{\hsize}{!}{\includegraphics[clip=true]{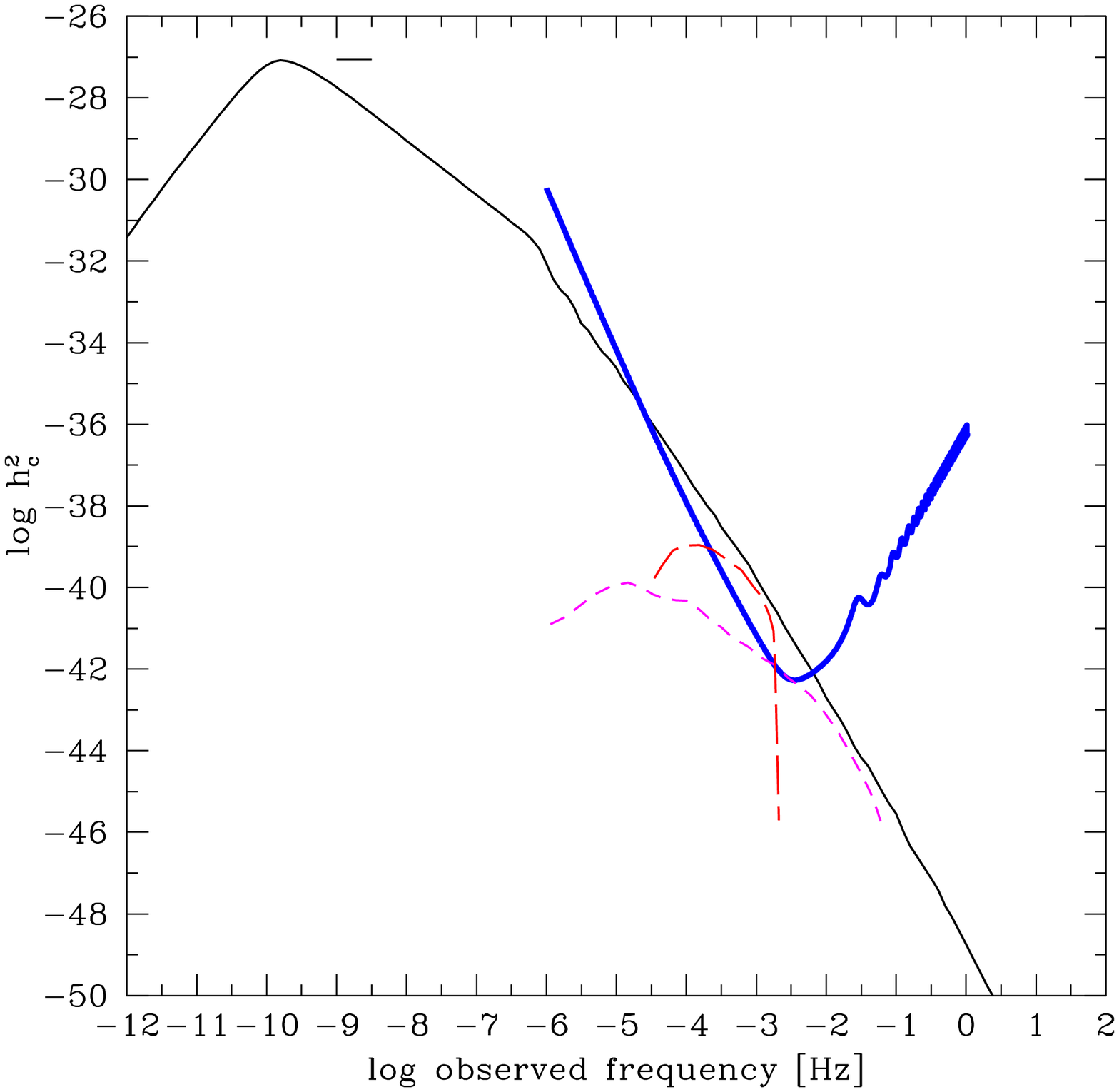}
   \includegraphics[clip=true]{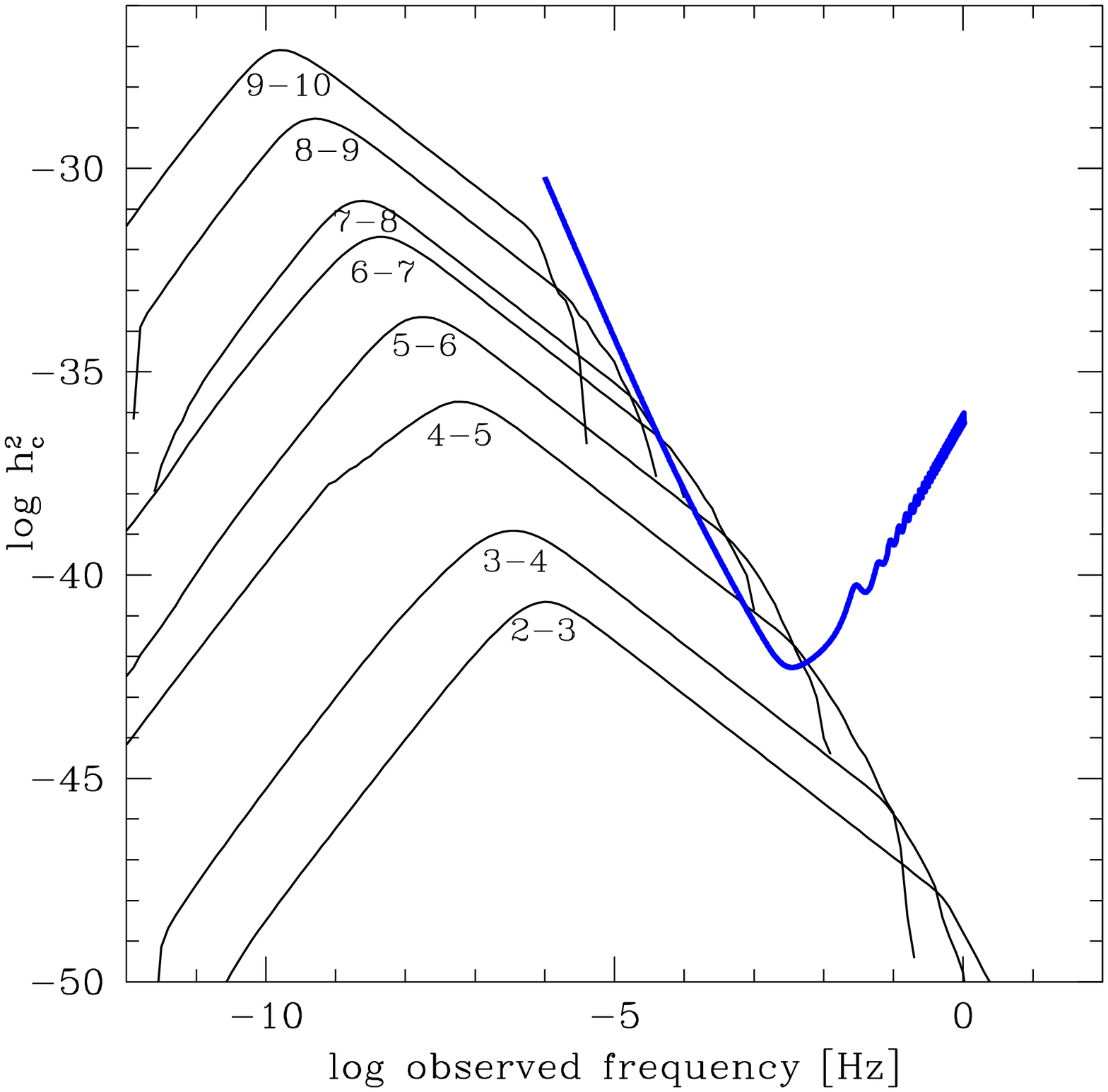}}
     \caption{Left panel: total GWB from inspiraling 
MBHBs. 
{\it Thin solid line}: square of the characteristic strain 
 vs. wave frequency. 
{\it Thick solid line}: {\it LISA} sensitivity curve.
{\it Short-dashed line}: expected strain 
from extragalactic WD-WD ~\cite{farmer}.
{\it Long-dashed line}: 
expected strain from unresolved galactic WD-WD ~\cite{nelemans}.
{\it Thick dash} at $f\simeq 10^{-9}$: current limits from pulsar 
timing experiments ~\cite{lommen}. 

Right panel: integrated GWB from inspiraling MBHBs  
in different mass ranges. From top to bottom (indicated by labels), 
the curves show the signal produced by events with $9<\log(m_1/\msun)<10\,$ , 
$8<\log(m_1/\msun)<9\,$,...,$2<\log(m_1/\msun)<3$.}  
        \label{Fig3}
    \end{figure}

Given the strain amplitude and the orbital evolution of the system, 
the energy spectrum integrated over the entire radiating lifetime 
$dE_{\textrm{gw}}/d\,\textrm{ln}f_r$ can be computed ~\cite{s1}. Then the 
total GWB due to the cosmological MBH binaries is calculated as:

\begin{equation}
c^2\rho_{\rm gw}(f)= {\pi \over 4}{c^2\over G} f^2h_c^2(f) =\int_0^{\infty} 
{dz\, {{N(z)}\over{1+z}}~{{dE_{\rm gw}} \over {d\ln{f_r}}}},
\label{pinnei}
\end{equation}
where $N(z)dz$ is the comoving number density of events in the 
redshift interval $z$, $z+dz$, and the factor $1/(1+z)$ accounts for the 
redshifting of gravitons ~\cite{pinnei}. 

The nature (stochastic or resolved) of the GWB can be assessed by counting
the number of events per resolution frequency bin whose observed signal
is above a given sensitivity threshold $h_{c,{\rm min}}$. 
The frequency resolution is simply $\Delta f=1/\tau$, where $\tau$ is the observation time, 
so the longer the observation, the smaller the noise. 
The GW signal due to a large population of sources 
is unresolved if there is, on average, 
at least one source per eight frequency resolution bins ~\cite{cornisc}. 

\section{Results and discussion: resolvable sources vs confusion noise}

   \begin{figure}
   \centering
   \includegraphics[width=4.0in]{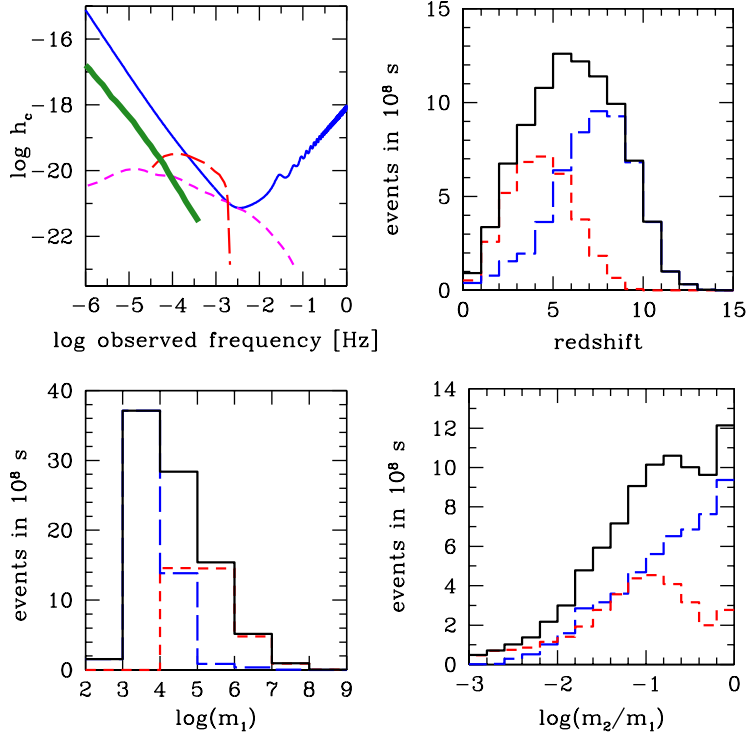}
      \caption{Upper left panel: confusion noise due to MBHBs  
as a function of frequency ({\it thick
solid line}). Other lines have the same meaning as in fig. \ref{Fig3}, 
left panel.
Upper right panel: differential redshift distribution of MBHBs resolved with 
$S/N>5$ by {\it LISA} in a 3-year mission ({\it solid lines}). 
The separate counts for MBs ({\it short-dashed line}) and IBs 
({\it long-dashed line}) are also shown. 
Lower left panel: mass distribution of the more massive memebr of 
MBHBs. Line style as in the right upper panel. 
Lower right panel: mass ratio distribution of MBHBs. Line
style as in right upper panel.  
                           }
         \label{Fig4}
   \end{figure}

The results for GWB are presented in fig. \ref{Fig3}.
Left panel shows the total background in terms of $h_c^2$ as a 
function of the observed frequency, while in the right panel the contribution
of binaries in different mass ranges is plotted. 
An extensive discussion is given in ~\cite{s1}.\\
{\it LISA} threshold is estimated combining the 
{\it LISA} single-arm Michelson sensitivity curve (taken from 
~\cite{simulator}) with the recent analysis 
of the {\it LISA} instrumental noise below $10^{-4}$ Hz 
~\cite{bender}.\\
    
Fig. \ref{Fig4}, upper left panel, shows the level of unresolved confusion noise
due to MBHBs, compared to the {\it LISA} instrumental noise curve and 
to the noise due to unresolved galactic and extragalactic WD-WD binaries. 
While MBHBs would be the most important sources
of astrophysical confusion noise below $10^{-4}$ Hz, the level is
more than an order of magnitude below the {\it LISA}
instrumental noise. Then, {\it LISA} will resolve the GWB due to MBHBs
into discrete sources.\\
The histogram in fig. \ref{Fig4}, upper right panel, shows  
the number of resolved sources per unit redshift above $S/N>5$, 
assuming a 3-year {\it LISA} observation. The total number of 
detectable systems is divided into ``merging'' binaries 
(MBs, i.e. events whose signal emitted at
the last stable orbit lies above the {\it LISA} sensitivity, see fig. 2)
and ``in-spiral'' binaries (IBs, i.e., events with integrated $S/N>5$, 
but with emission amplitude at the last stable orbit well below the {\it LISA}
sensitivity). {\it LISA} will resolve about 90
events in a 3-year mission ($\approx$ 35 
MBs and $\approx$ 55 IBs) up to $z\simeq12$.
Finally, lower left and right panels show the number of sources as a 
function of $m_1$ and of $m_2/m_1$ respectively. 
While we expect observable MBs to involve
systems of about $10^5-10^6\msun$ with mass ratios of the order of 0.1-0.2,
IBs have lighter masses and mass ratios near unity. 

\section{Conclusions}
We have computed the GW signal (in terms of the characteristic 
strain spectrum) from the cosmological population of inspiraling MBHBs  
predicted to form 
at the center of galaxies in a hierarchical structure formation scenario. 
We note here that, while any hierarchical clustering model in $\Lambda$CDM
cosmology gives similar halo merging rates (e.g,~\cite{loeb}), the coalescence time of MBHBs is a matter of
(more uncertain) estimates of the dynamical friction and hardening time scales. 
On the other hand, as long as the coalescence time is short compared to the Hubble time, 
we expect that our results are only weakly dependent on the details of MBH dynamics.\\
In the {\it LISA} window ($10^{-5.5}\leq f\leq 0.1$ Hz), 
the main sources of GWs are MBHBs in the mass range $10^{3}\leq m_1\leq 
10^{7}$ $\msun$. With a plausible lifetime of $\simeq 3$ years, {\it LISA} will resolve the GWB 
into $\approx 90$ discrete sources above $S/N=5$ confidence level. 
Among these, $\approx 35$ are MBs, the other are IBs.  
Most of the observable MBs are at $3 \leq z \leq 7$, while
IBs can be detected up to $z\approx12$.
Once subtracted resolvable sources to the total GWB, 
the remaining confusion noise level
is expected to be well below the {\it LISA} sensitivity threshold ~\cite{s2}.\\
While {\it LISA} will make it possible to probe the coalescence of 
early black hole binaries in the universe, it may be not be able
to observe the formation epochs of first MBHs. We conclude remarking 
that the bulk of detections involves binaries in the range $10^3-10^5\msun$,
a range where black holes have never been observed.

\ack{We thank A. Vecchio and P. Bender for helpful discussions.}

\end{document}